# Revisiting the complex nuclear region of NGC 6240 with *Chandra*

G. Fabbiano[1], A. Paggi[2, 3, 4], M. Karovska[1], M. Elvis[1], E. Nardini[5, 6], Junfeng Wang[7]

1. *Center for Astrophysics | Harvard & Smithsonian, 60 Garden St. Cambridge MA 02138, USA*
2. *INAF-Osservatorio Astrofisico di Torino, via Osservatorio 20, 10025 Pino Torinese, Italy*
3. *INAF-Osservatorio Astrofisico di Torino, via Osservatorio 20, 10025 Pino Torinese, Italy*
4. *Istituto Nazionale di Fisica Nucleare, Sezione di Torino, I-10125 Torino, Italy.*
5. *Dipartimento di Fisica e Astronomia, Università di Firenze, via G. Sansone 1, I-50019 Sesto Fiorentino, Firenze, Italy*
6. *INAF—Osservatorio Astrofisico di Arcetri, Largo Enrico Fermi 5, I-50125 Firenze, Italy*
7. *Department of Astronomy, Physics Building, Xiamen University, Xiamen, Fujian, 361005, China*

Abstract

We present a reanalysis of the cumulative ACIS S *Chandra* data set pointed at the double AGNs of the NGC 6240 merging galaxy, focusing on the hard energy bands containing the hard spectral continuum (5.5-5.9 keV), the redshifted Fe I K$\alpha$ line (6.0-6.4 keV), and the redshifted Fe XXV line (6.4-6.7 keV). We have used to the full the *Chandra*' telescope angular resolution, and we have modeled the *Chandra* PSF by comparing pre-flight calibration model to the data for the two bright AGNs. With two complementary approaches: (1) studying the residuals after PSF subtraction, and (2) producing reconstructed Expectation through Markov Chain Monte Carlo (EMC2) images, we are able to resolve structures extending from ~1 kpc to <200 pc in the S AGN. The latter are within the sphere of influence of this BH. We find significant extended emission in both continuum and Fe lines in the ~2" (~1 kpc) region surrounding the nuclei, in the region between the N and S AGN, and in a sector of PA 120-210 deg. extending to the SE from the centroid of the S AGN surface brightness. The extended Fe I K$\alpha$ emission is likely to originate from fluorescence of X-ray photons interacting with dense molecular clouds, providing a complementary view to recent high-resolution *ALMA* studies. The non-thermal emission is more prevalent in the region in between the two active X-ray nuclei, and in the N AGN. We do not find strong evidence of X-ray emission associated with the 3rd nucleus recently proposed for NGC 6240.

## 1. Introduction

The highly disturbed merger galaxy NGC 6240 (z=0.02448; D~111 Mpc; 1"~511 pc, NED[1]) is well known to host two highly absorbed Compton-thick active galactic nuclei (CT AGNs), with ~0.8 kpc separation. These CT AGNs were discovered in X-rays with *Chandra* (Komossa et al. 2003), and later identified as radio sources (e.g., Gallimore & Beswick 2004).

---
[1] http://ned.ipac.caltech.edu/

Subsequent CO (1-0) observations of this nuclear region led to the discovery of high velocity outflows (Feruglio et al. 2013). *Chandra* observations provided evidence of the X-ray counterpart of these outflows, the hard, thermal X-ray emission of the interstellar medium (ISM) shocked by these winds (Feruglio et al. 2013; Wang et al. 2014). While AGN outflows may be responsible for these shocks (Feruglio et al. 2013), the intense star formation of the circumnuclear region, and the similarity of the extended Fe XXV spatial distribution with that of molecular hydrogen in the region, also suggested stellar/supernova winds as a probable cause (Wang et al. 2014).

More recent multi-wavelength, high angular resolution, observations of NGC 6240 have uncovered additional complexity in this nuclear region. A combined *MUSE* and *HST* study has provided a high spatial resolution characterization of the ionization properties, and has led to the possible discovery of a third nucleus and to stellar velocity dispersion mass estimates for the black holes of both CT AGNs and the third nucleus: (~4, ~7) x $10^8$ Msol for the N and S AGN; ~9 x $10^7$ Msol for the third nucleus (Kollatschny et al. 2020). High resolution (0.03'' ~15 pc) *ALMA* observations have revealed a concentration of CO (2-1) emitting clouds in a clumpy turbulent stream between the two CT AGNs with a total molecular cloud mass of ~9 x $10^9$ Msol (Treister et al. 2020).

These recent results have spurred us to take a new look at the *Chandra* ACIS-S observations of NGC 6240 used in the Wang et al. (2014) study. Our re-analysis is also informed by what we have learned in the intervening years about the spatial and spectral properties of the hard (>3 keV) emission of nearby CT AGNs.

As is generally the case for CT AGNs (e.g., Levenson et al., 2002), in NGC 6240 the Fe I K$\alpha$ (6.4 keV) line is prominently associated with the two nuclei (Komossa et al. 2003, Wang et al 2014). This nuclear 6.4 keV line is generally believed to be due to fluorescence induced by the AGN photons interacting with the dusty 'torus' surrounding the Compton thick nuclei (e.g. Baloković et al., 2018). Although Wang et al. 2014 reported possible extended emission in the Fe I K$\alpha$ (6.4 keV) line, they concentrated on the spatial properties of the hard continuum and the Fe XXV (6.7 keV) emission line in NGC 6240, both of which were found to be convincingly extended, and related them to the thermal emission of the shocked ISM.

However, more recent high-resolution *Chandra* studies of several CT AGNs, have demonstrated that the 6.4 keV line, and the associated non-thermal hard continuum emission, may have extended components with sizes ranging from several 100 pc to a few kpc. These extended components can be responsible for up to ~20-30% of the total line emission (Marinucci et al. 2012, 2017; Bauer et al. 2015; Koss et al. 2015; Fabbiano et al 2017, 2018a, b; Jones et al. 2020; Ma et al. 2020a). These observations of spatially extended 6.4 keV line and hard continuum emission suggest that these components are not restricted to the interactions of AGN photons with the nuclear torus (~0.1 pc or less from the AGN; Suganuma et al., 2006) but may also arise from interactions with dense molecular clouds in the disk or other structures of the host galaxy (Fabbiano et al. 2018b, 2019a). A direct example of such interaction is provided by a recent *Chandra* study (Fabbiano et al. 2018c), where 6.4 keV Fe I K$\alpha$ emission was spatially associated with the CO (2-1) ~26 pc circum-nuclear rotating disk of NGC 5643 discovered with *ALMA* by Alonso-Herrero et al. (2018).

In this paper, we revisit the spatial behavior of the emission in the area immediately surrounding and in-between the active nuclei of NGC 6240, to explore the relative amount of thermal and non-thermal emission in the inner circum-nuclear region, and investigate a possible association of the Fe I K$\alpha$ line with the molecular cloud distribution (Treister et al 2020). Exploiting the angular resolution of *Chandra* fully, we also explore a possible X-ray counterpart to the third nucleus suggested by Kollatschny et al. (2020) to the SE of the southern CT AGN. We have reanalyzed the archival *Chandra* data, to obtain the most accurate estimate of the hard emission (continuum and lines) near the CT AGNs of NGC 6240. We have taken particular care in preparing the data and in modeling the *Chandra* Point Spread Function (PSF), as explained in Section 2. The results of this study are reported in Section 3 and discussed in Section 4. Our main conclusions are summarized in Section 5.

## 2. Data Preparation

We have used the same data set as in Wang et al. (2014), which includes all the ACIS-S observations with NGC 6240 at the aimpoint. It consists of two imaging ACIS-S observations (ObsID 1590, 12713) with a total exposure time of 183 ks, and of two ACIS-S HETG grating observations (ObsID 6908, 6909) with a total exposure time of 302 ks. We use the 0th order images of the grating observations. Given the response of the gratings, the combined effective ACIS-S imaging exposure time in the hard >5 keV band we use in this study is 363 ks, as in Wang et al. (2014).

In this study, we propose to exploit the full resolution of *Chandra*, to peer in the region near the two CT AGNs. For this reason, it is important to make sure that the individual ACIS observations are merged as precisely as possible. Also, it is important that we use the most appropriate PSF to model the unresolved AGN surface brightness. The presence of two readily detectable AGNs in the nuclear region of NGC 6240 has helped in both tasks, as described below.

For the analysis, we made use of CIAO[2] and DS9[3] both for image display and as a CIAO interface. To optimize the resolution, we used subpixel binning of 1/8 and 1/16 of the 0".492[4] ACIS instrumental pixel. This method has been validated by several works, including our own (see e.g. Harris et al. 2004; Siemiginowska et al. 2007; Karovska et al. 2010; Wang et al. 2011a, b, c; Paggi et al.2012; Fabbiano et al. 2018b). It is conceptually comparable to the *HST* drizzle imaging (Fruchter and Hook, 2001) and exploits the sharp central peak of the *Chandra* PSF and the well-characterized *Chandra* dither motion[5] to retrieve the full *Chandra* mirror resolution.

### 2.1 Merged Data Set

To obtain a fine alignment of the nuclear emission, we registered the observations using the centroids of the images in the energy band 6.0-6.4 keV, where the nuclear emission is dominated by the prominent 6.4 keV Fe I K$\alpha$ line, which has a strong point-like component (see Wang et al.

---

[2] https://cxc.harvard.edu/ciao/
[3] https://sites.google.com/cfa.harvard.edu/saoimageds9
[4] https://cxc.harvard.edu/proposer/POG/html/chap6.html#tab:acis_char
[5] https://cxc.harvard.edu/proposer/POG/html/chap5.html#tth_sEc5.3

2014); this line is observed at 6.2 keV, as expected given the redshift of NGC 6240, z=0.02448. Visual inspection of the images led us to choose the N AGN as a reference, because the Fe I K$\alpha$ image looks symmetrically round, as expected from point-like emission. To evaluate the centroid of the nuclear source for each observation, we used the following procedure. First, we produced 6.0-6.4 keV band images with sub pixel scale of 1/8 of the native ACIS pixel; second, we smoothed these images with a two-dimensional Gaussian kernel with $\sigma = 2$ image pixels, and selected the brightest pixel in the resulting images; finally, using the original 1/8 pixel image for each ObsID, we found the centroid of the count distribution within a circular region with a radius of 0".2 centered on the brightest pixel of the corresponding smoothed image.

We then used the *wcs_update* CIAO task to match the N AGN centroids, using the deepest observation (i.e., OBSID 12713) as a reference. The shifts needed to match the centroids range between 0".15 and 0".23, compatible with the 0.8'' *Chandra* absolute astrometric uncertainty[6]. Figure 1 shows the results of the registration procedure. We reprojected the shifted observations to a common tangent point with the *reproject_obs* task and merged the reprojected observation with the *flux_obs* task.

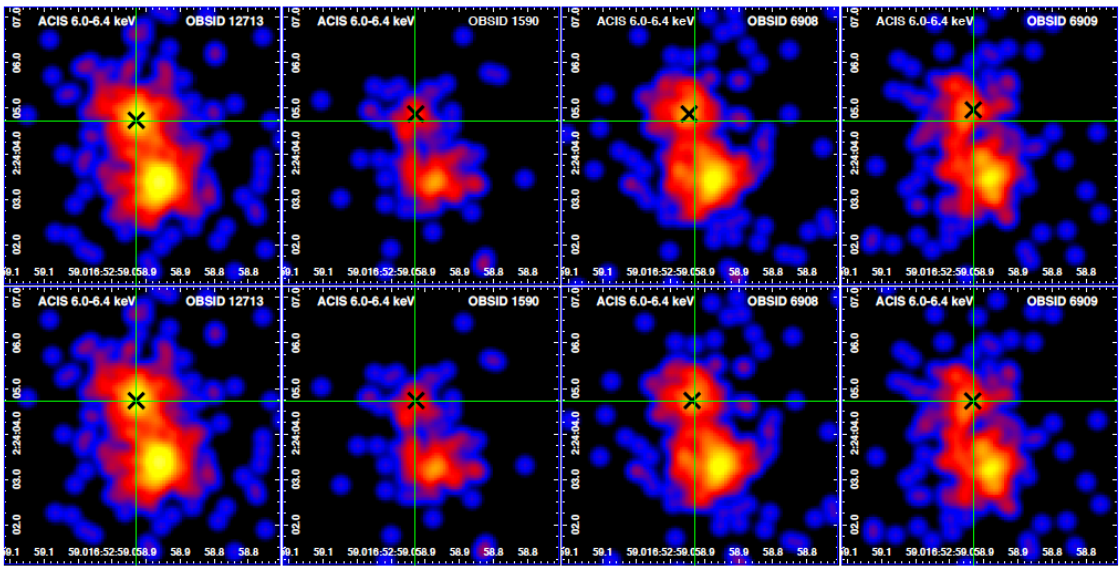

**Figure 1** – Neutral iron (6.0-6.4 keV) band images for the *Chandra* observations of the nuclear region of NGC 6240 with a subpixel of 1/8 and 2 image pixel Gaussian smoothing. The black crosses mark the position of the centroids. The upper and lower panels show the images for the observations before and after the centroid registration. For all the panels, N is to the top and E to the left.

Figure 2 shows on the left the 6.0-6.4 keV merged images obtained from the observations performed in pure imaging mode (OBSIDs 12713 and 1590), and on the right the full data set adding to these the 0$^{th}$ order images of the HETG grating observations (OBSIDs 6908 and 6909).

---

[6] https://cxc.cfa.harvard.edu/cal/ASPECT/celmon/

The latter add to the signal to noise, without no change in quality. Both images suggest an extension of the surface brightness of the S AGN to the SE, which is confirmed by our following analysis. In what follows, we use the full available dataset.

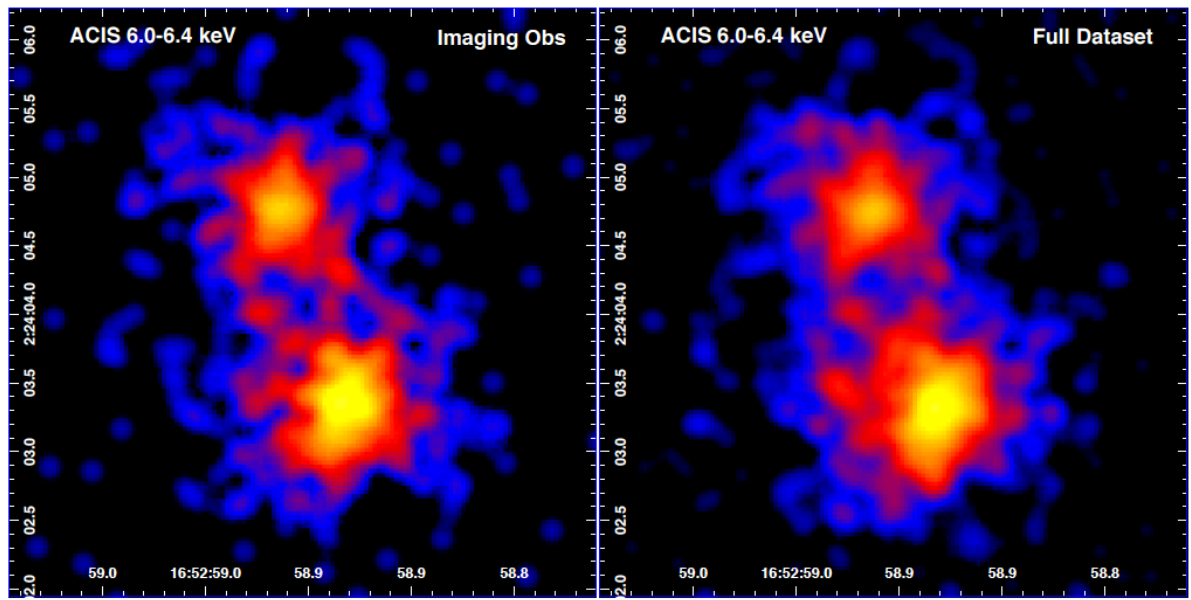

**Figure 2** – Neutral iron (6.0-6.4 keV) band images with subpixel 1/16 and 5 image pixel Gaussian smoothing, obtained merging the imaging observations (left) and the full dataset (right). For both panels, N is to the top and E to the left.

**2.2 Point Spread Function (PSF) modeling**

The *Chandra* mirror PSF was well-calibrated before launch and characterized as function of photon energy[7]. This PSF is well reproduced in flight by HRC observations, except for a small PSF artifact situated at a radial distance of ~0".6 – 0".8 from the centroid of the PSF, in a well determined angular position that depends on the roll angle of the observation. This artifact contains ~6% of the counts in the PSF, and therefore it is important to consider near strong point sources (Juda & Karovska 2010[8]; see also the CIAO documentation page[9]).

However, we are using ACIS data in sub-pixel binning for our work, for which in-flight calibrations of the PSF have not yet been developed by the CXC. Given that we are attempting to investigate sub-arcsecond scale features, it is important to carefully model the PSF. Fortunately, our data include two prominent point-like sources from the two AGNs, which we could exploit as empirical observations of the PSF. With our observations of the N and S AGN we can then effectively measure the PSF, and compare it to simulations based on the pre-flight calibrations. We performed model-data comparisons using both 1/8 and 1/16 subpixel binning, with consistent results. We also compared separately the PSF models with the imaging observations (ObsID 1590 and 12713) and the 0$^{th}$ order grating observations (6908, 6909) data, to exclude any possible unknown effect in the 0$^{th}$ order grating PSF. We obtained consistent results as for the ACIS S imaging observations.

---

[7] https://cxc.harvard.edu/proposer/POG/html/chap4.html#tth_sEc4.2.3
[8] http://hea-www.harvard.edu/~juda/memos/HEAD2010/HEAD2010_poster.html
[9] https://cxc.cfa.harvard.edu/ciao/caveats/psf_artifact.html

The *Chandra* PSF was simulated using rays produced by the *Chandra Ray Tracer* (*ChaRT*[10]) projected on the image plane by *MARX*[11]. For each observation we generated the average from 1000 PSF simulations centered at the coordinates of the centroids shown in the lower panels of Figure 1. We then produced images of these PSFs in each energy band used in this study, which correspond to those in Wang et al. (2014): 5.5-5.9 keV, 6.0-6.4 keV, and 6.4-6.7 keV. These energy bands are representative of the hard continuum emission, and the red-shifted Fe I K$\alpha$ I and Fe XXV emission lines, respectively. The PSFs were then merged to be compared with the merged data set.

MARX allows the application of an 'aspect blur' to the simulated PSF, to account for possible smearing of the image resulting from imperfect photon positioning. We found that the recommended 0".25 blur[12] created a PSF that did not match the central regions of the observed AGNs surface brightness. We investigated the amount of blur to be applied in two ways: we derived an encircled energy curve from the strong point-like Fe I K$\alpha$ emission line (observed in the 6.0-6.4 keV band) and compared it with the pre-launch calibrations, within a radius of 0".4; we also derived a set of model PSFs for a range of blur parameters, and compared them with the radial profile of the Fe I K$\alpha$ sources. In both cases, the radial profile of the observed emission is consistent with the *Chandra* PSF in the same energy band when no blur is applied. In particular, the comparison of the X-ray surface brightness with the PSF shows that a central region of positive residuals begins to appear for blur factor 0.04 [see e.g., Figure 3; Left and Center panels; The S AGN surface brightness excludes the SE excess, as in Figure 5 below]. We therefore did not apply a blur factor to the model PSF.

We explored empirically how to best match the model PSF to each merged AGN observation, on the assumption that the observed emission of each AGN consists of an unresolved component that should match the PSF, plus possible residual extended emission. We explored matching the PSF to the data for radii from the data/PSF centroid, ranging from 0".2 to 0".5, taking into account the contribution of the PSF wings to each other's normalization circle. For higher enough signal to noise ratios, it would be preferable to use as small a radius as possible to normalize the PSF to the data. However, the statistical uncertainties in the inner bins are large enough to bias the fit, when the nominal values are used, resulting in regions of cumulative negative residuals, i.e. over-subtracting the PSF (Figure 3; Right panel, for the N nucleus; a similar behavior is observed for the S nucleus, excluding the 'excess' region, see below).

---

[10] http://cxc.harvard.edu/ciao/PSFs/chart2/
[11] http://space.mit.edu/CXC/MARX/
[12] https://cxc.cfa.harvard.edu/ciao/why/aspectblur.html

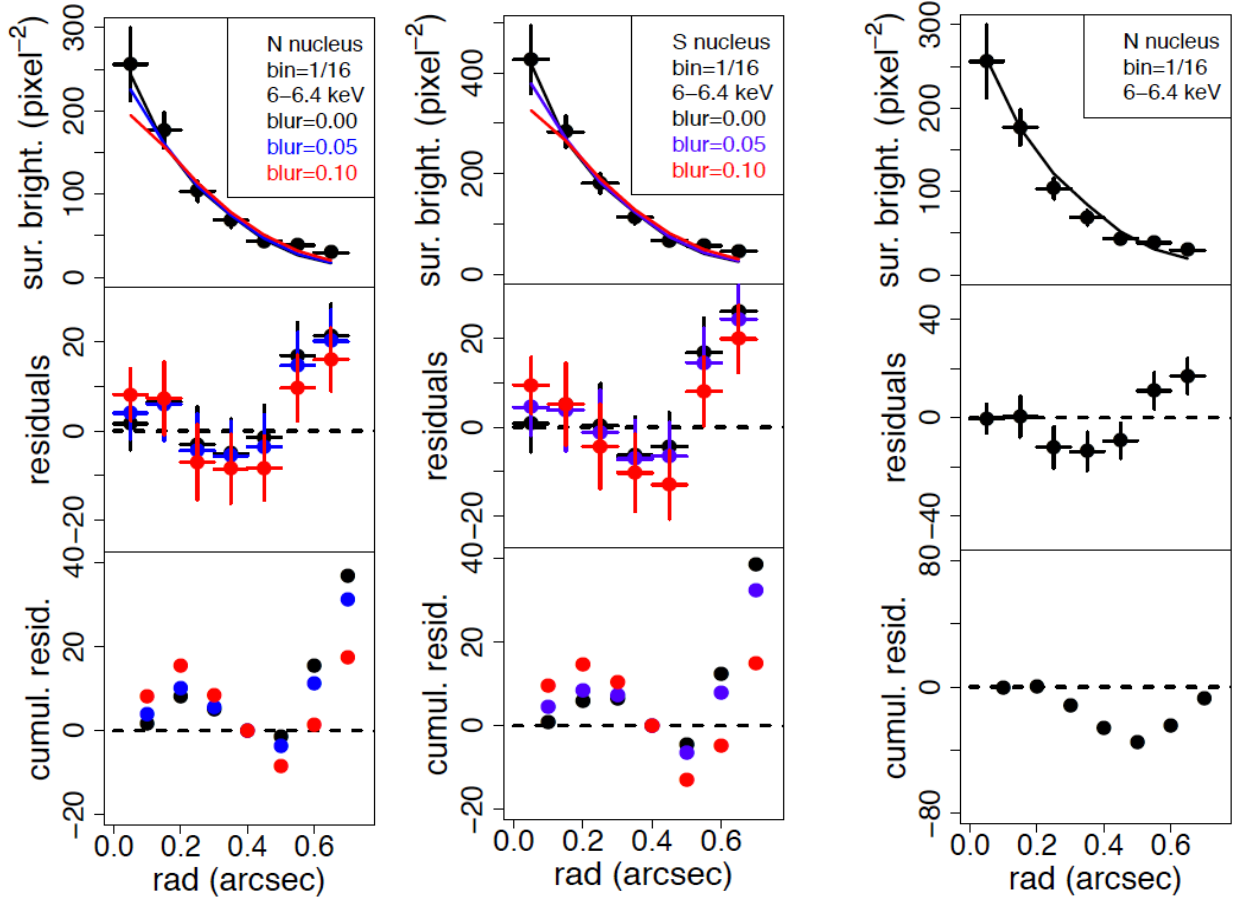

**Figure 3** – <u>First two panels from left</u>: Observed 6.0-6.4 keV (Fe I Kα) surface brightness (in counts/pixel$^2$), residual counts from the PSF fit, and cumulative residual counts within circles of given radii of the surface brightness for the N AGN (left) and S AGN (center), for PSFs with increasing blur parameter (as labeled). The PSFs were all normalized within 0".4 from the centroid. <u>Right panel</u>: Observed surface brightness (in counts/pixel$^2$), residual counts from the PSF fit, and cumulative residual counts within circles of given radii, for the N AGN, 6.0-6.4 keV (Fe I Kα). The PSF was normalized within 0".2 of the centroid, resulting in a region of prominent negative residuals (i.e. PSF larger than the data at a given radius).

For the N AGN, normalizing the model PSF to the data within 0".4, we find an excellent agreement of the profiles with the PSF within 0".5. The cumulative residual counts within 0".5 is consistent with zero. At larger radii, excess extended emission is apparent. Figure 4 shows the radial profiles of the data and PSF for the N AGN for 1/16 binning for the data. In all our experiments, we obtain consistent results using 1/8 and 1/16 binning.

In the S AGN, instead, the match between PSF and data was not straightforward, because of excess emission within 0".4 from the nuclear centroid in the SE quadrant. Inspection of the radial surface brightness profiles at different azimuthal angles showed that this excess emission is concentrated in the region at Position Angle (PA) between 120-210 degrees (counterclockwise from N).

Excluding this azimuthal sector, we were able to match the PSF well to the remaining data within 0".5, normalizing the PSF to the data within a radius of 0"4. Figure 5 shows the azimuthal sector of the S AGN emission, which we used to match the PSF. Figure 6 shows the excess sector, also with 1/16 binning. We obtain consistent results using 1/8 binning.

We conclude that the merged PSFs with no blur we have produced are excellent models of the merged unresolved emission of the NGC 6240 AGNs within 0".5 of the centroids, when using 1/8 and 1/16 subpixel data.

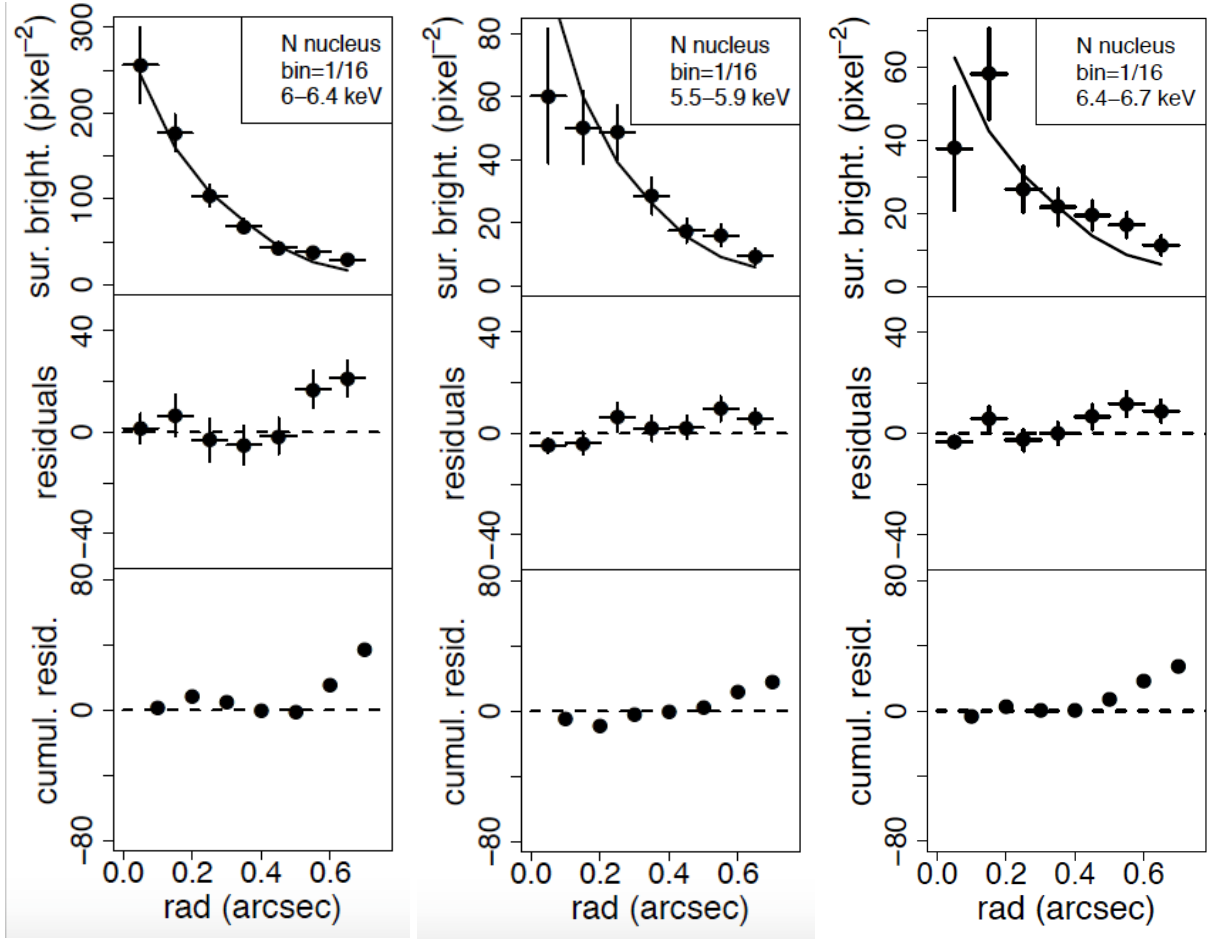

**Figure 4** – Observed surface brightness (in counts/pixel$^2$), residual counts from the PSF fit, and cumulative residual counts within circles of given radii, for the N AGN. Left: 6.0-6.4 keV (Fe I Kα; Middle: 5.5-5.9 keV continuum; Right, 6.4-6.7 keV, containing Fe XXV. The PSF was normalized within a radius of 0".4.

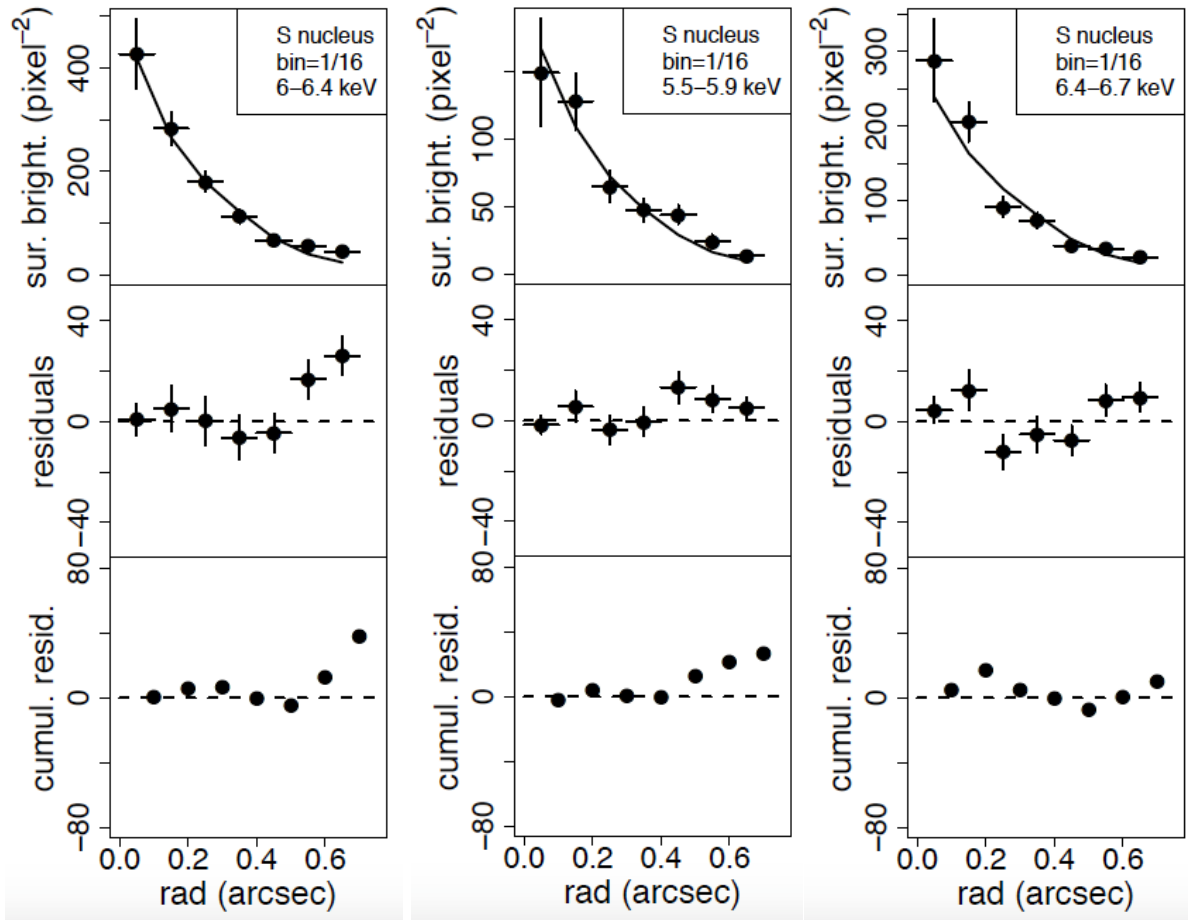

**Figure 5** - PSF-data comparisons for the S AGN, for the azimuthal region of good PSF-data agreement (see text). The three panels within each figure are analogous to those in Fig. 4.

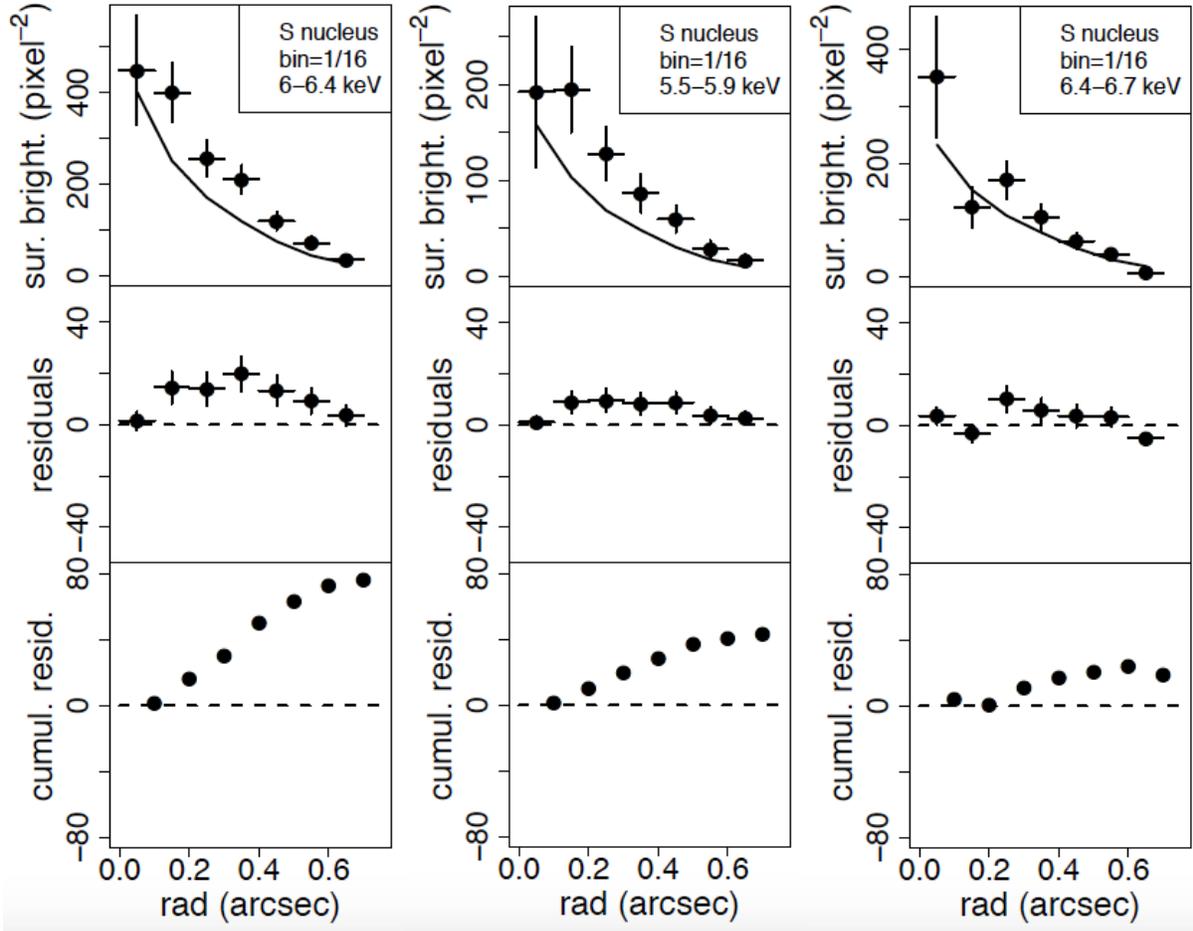

**Figure 6** - PSF-data comparisons for the S AGN, for the 'excess' region (PA 120-210 degrees, counterclockwise from N) using the same PSF matching as in Fig. 5. The three panels within each figure are analogous to those in Fig. 4.

## 3. Analysis of the Circumnuclear Surface Brightness and Results

Using the merged data set and our model PSFs for the two nuclear AGNs discussed in Section 2, we have investigated the spatial properties of the circumnuclear emission of NGC 6240 in three energy bands: 5.5-5.9 keV, where the emission is a featureless hard continuum; 6.0-6.4 keV, dominated by the Fe I K$\alpha$ 6.4 keV line in the observed frame; and 6.4-6.7 keV, where the Fe XXV K$\alpha$ line is observed. These bands correspond to the spectral regions illustrated in Figures 4-6.

We have taken two approaches to this analysis. First, we analyze directly the residual images obtained in the three energy bands, after subtracting the normalized PSFs. This approach highlights directly the features of the diffuse emission and allows the extraction of counts and statistical evaluation of different features in a straightforward way. Then, we have applied Markov Chain Monte Carlo techniques to the original, un-subtracted images, to obtain a reconstruction of the original emitted surface brightness before the spreading of photons introduced by the *Chandra* PSF. While this reconstruction in principle would give the 'truest' representation of the emission region, it is dependent on the PSF model adopted, and will reflect any systematic uncertainty in this model.

### 3.1 PSF-Subtracted Images of the Nuclear Region

In Section 2.2 we have shown that comparisons between the observed surface brightness of the two circumnuclear regions with the expected PSF reveals two types of discrepancies: (1) an excess surface brightness at larger radii than 0".5, and (2) a localized excess in the S AGN in the 120-210 degrees PA azimuthal region SE of the nuclear centroid. Here we explore these features and their spatial distribution in detail. We also address possible biases introduced by the PSF artifact in these images.

Figure 7 compares the residuals after PSF subtraction for the central ~1 kpc (~2") radius region of NGC 6240 with 1/8 binning, in the three energy bands under study. Only positive residuals are shown. Some negative pixels occur within the central 0''.4 circles, consistent with statistical noise. As shown in Figures 4 and 5 (Section 2), the cumulative residual counts are consistent with zero within the circles, with the exception of the 'exclusion region' to the SE of the S AGN (Figure 6). To better visualize the extended emission, we have produced the adaptively smoothed residual maps shown in Figure 8 with the *CIAO* tool *dmimgadapt*, were we used 1/16 binned data smoothed with Gaussians, with 9 counts under the kernel and sizes ranging from ½ to 15 image pixels, in 30 logarithmic steps.

Although there is significant excess emission in all three energy bands, the excess is particularly significant in the Fe I K$\alpha$ band. In both this image and in the Fe XXV image there is a concentration of residual counts in the area between the two AGNs, which is not seen in the 5.5-5.9 keV continuum image. In the Fe I K$\alpha$ (top) of Figure 7 we also show the regions affected by the PSF artifact, for each observation in our data set (red: ObsID 1590; cyan: 6908; magenta: 6909; yellow: 12713). The overall artifact region shape is also reproduced in Figure 8. This artifact results in a spurious positioning of 6% of the counts of the nuclear AGN in each of these regions. Since the two nuclear sources are bright, it is important to assess the impact of this effect. We estimated the PSF artifact counts using the model PSFs for each nucleus.

Table 1 lists the cumulative counts in the residual images within a 2" radius (~1 kpc). In parentheses, we give the cumulative counts expected from the PSF artifact within a 2" radius. Given the total number of extended counts, this effect is small (~20%), and does not change our conclusions on the presence of extended circum-nuclear emission. However, there may be localized effects, since the artifact areas are concentrated in the regions between the two nuclei and to the NW of the northern nucleus. Extraction of counts from these regions indeed shows values consistent within statistics with those expected from the PSF artifacts. To investigate further the presence of diffuse emission in the larger region between the two AGNs, we extracted counts from the box highlighted with a dashed line in the top panel of Figure 8. Subtracting the counts expected from the PSF artifact within this box, we find that there is highly significant (5.6 $\sigma$) diffuse emission in this region in the Fe I K$\alpha$ band and a 3 $\sigma$ excess in the Fe XXV band. No significant excess is found in the 5.0-5.5 keV hard continuum. The excess counts in the box are listed in Table 1.

As discussed in Section 2.2.1, the N AGN surface brightness is well fitted with the PSF, and correspondingly the residuals in the 0".4 N circle are consistent with zero, as shown in Table 1. In the S AGN instead, we find significant count excess within the 0".4 circle in all the energy bands, but most significant in the Fe I K$\alpha$ band (Table 1). This excess originates from the 120-210 degrees PA sector SE of the nucleus. Outside this sector, the surface brightness distribution of the S AGN is well fitted by the PSF (see Figure 5). These circumnuclear circles of 0".4 radius (~200 pc) roughly correspond to the spheres of influence of each nuclear Black Hole (BH), i.e. the region within which the BH contributes ½ of the total mass (Medling et al. 2015).

The excess to the SE of the S AGN is free of contamination from the PSF artifact. This excess is most prominent in the Fe I K$\alpha$ emission line. It extends outwards in SE direction to ~1''.5 radius (Figure 8). Figure 8 suggests that this feature may fan out to a wider angle outside the 0".4 radius circle, where it occupies the 120-210 degrees sector. Using the image for guidance, we extracted the counts contributed by this feature, from 110-210 degrees PA angular sectors of 1".5 radius centered on the Fe I K$\alpha$ centroid of the southern nucleus. This region is shown in red in the 6.0-6.4 keV panel of Figure 8. We list in Table 1 the total counts estimated for this SE feature in each energy band.

**Table 1 – Cumulative Extranuclear Residual Counts**

| Energy | 2'' circle (PSF art.) | Box Excess | N - 0".4 circle | S - 0".4 circle | SE feature |
|---|---|---|---|---|---|
| 5.5 - 5.9 keV | 181.5 ± 13.5 (30) | 11.0 ± 5.4 | −2.3 ± 1.5 | 28.6 ± 5.3 | 66.2 ± 8.2 |
| 6.0 - 6.4 keV | 346.1 ± 18.6 (71) | 53.7 ± 9.6 | 1.6 ± 1.3 | 47.8 ± 6.9 | 122.9 ± 11.1 |
| 6.4 - 6.7 keV | 151.7 ± 12.3 (36) | 25.7 ± 7.1 | 0.6 ± 0.7 | 17.5 ± 4.2 | 27.2 ± 5.2 |

Notes:
1. The errors are statistical 1 $\sigma$.
2. The local background counts in a 2" radius circle is ~5, 18, 7 counts in order of increasing energy, and <1 count in each of the nuclear circles.
3. The numbers in parentheses in the 2" circle are the total counts expected from the PSF artifact.
4. Box excess gives the residual counts in the region between the two nuclei, within the box in Figure 8, once the PSF artifact contribution has been subtracted

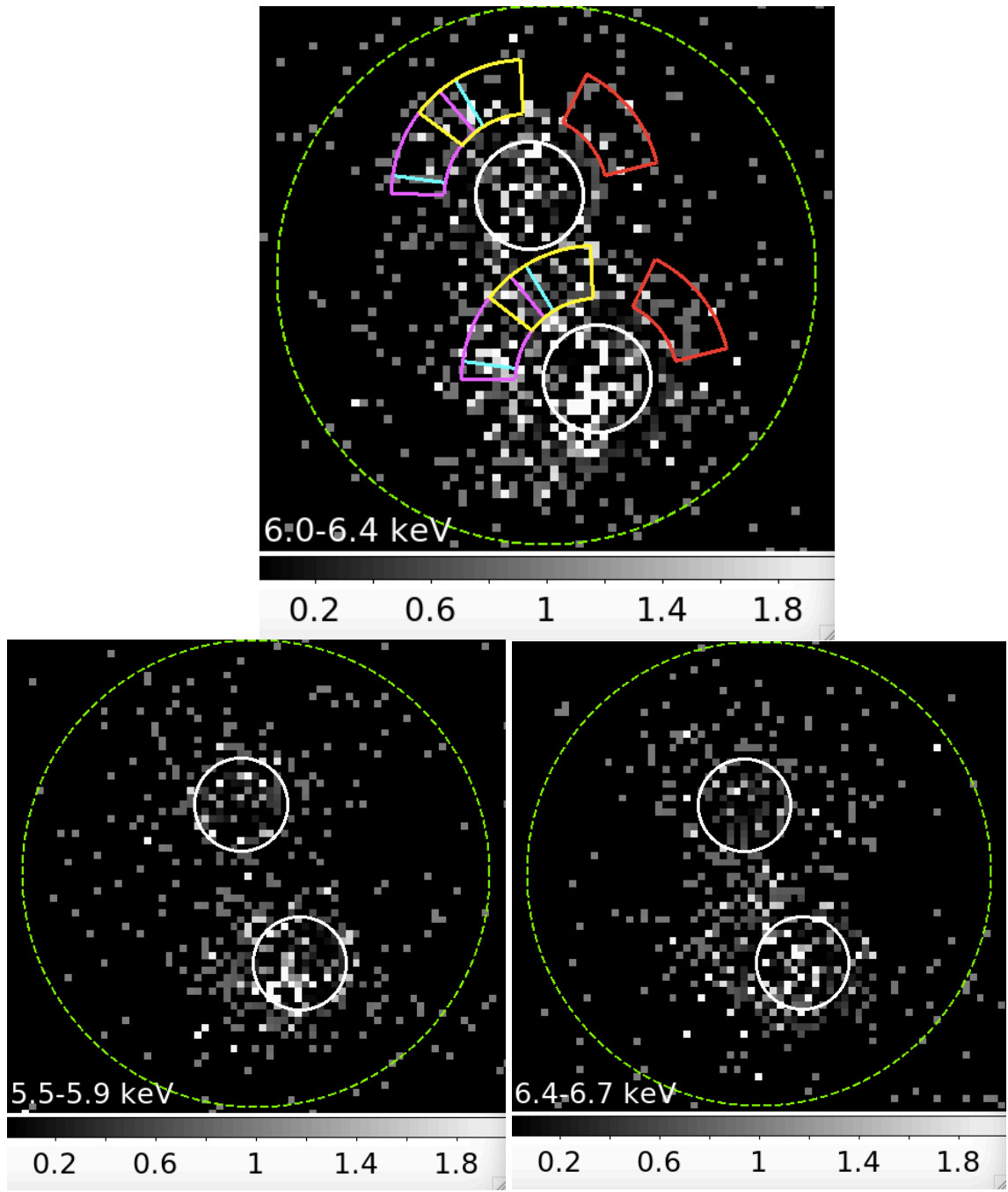

**Figure 7** – 1/8 binning images of positive residuals after PSF subtraction (see text), in three energy bands: 6.0-6.4 keV, Fe I Kα line (top); 5.5-5.9 keV, continuum (bottom-left); 6.4-6.7 keV, Fe XXV (bottom-right). N is to the top and E to the left. The scale is linear. The larger, green, circle has a radius of 2'' (~1 kpc). The two smaller circles have radii of 0''.4 (~200 pc) and correspond to the areas dominated by the AGN point-like emission. The regions affected by the PSF artifact that augments the flux by 6% of the dominant PSF are shown in the top panel with colored annular sectors, for each observation (red: ObsID 1590; cyan: 6908; magenta: 6909; yellow: 12713). The same areas apply to all the energy bands shown.

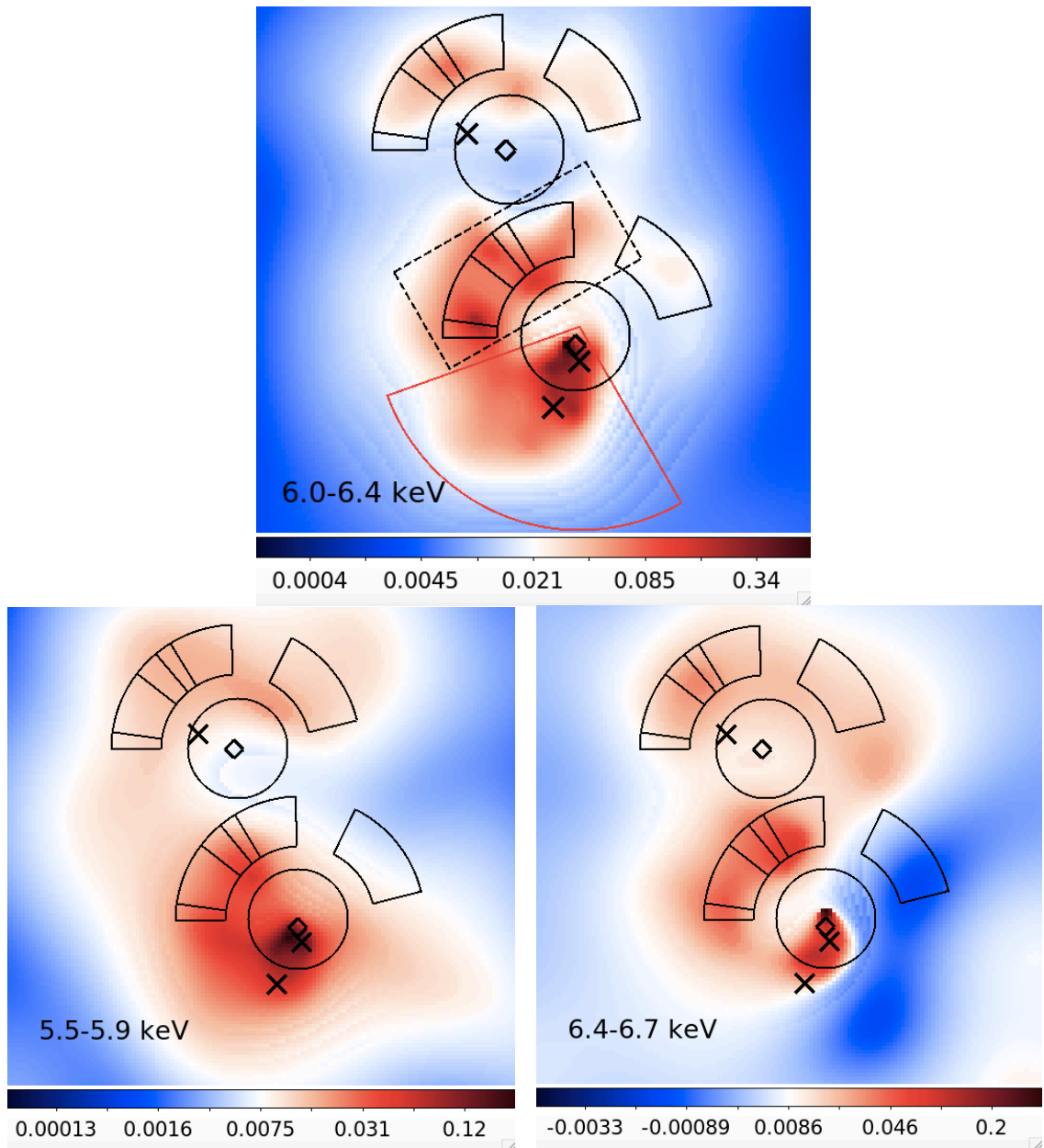

**Figure 8** – Adaptively smoothed images of the PSF-subtracted residuals in inner region of NGC 6240, in three energy bands, to visualize the SE extended feature connected with the southern AGN. This feature is more prominent in the Fe I Kα band (upper panel) but is also detected in the hard continuum emission (lower-left panel) and in Fe XXV (lower right panel). The circles are the 0".4 (~200 pc) radius regions used to normalize the model PSF to the data and the circular annuli are the regions contaminated by the PSF artifact (see text). N is to the top and E to the left in all the panels. The color scale is logarithmic in units of counts per image pixel (1/16 of the instrumental pixel). The diamonds represent the positions of the radio nuclei (Gallimore & Beswick 2004), while the crosses are at the position of the three optical nuclei (Kollatschny et al. 2020). The dashed box in the 6.0-6.4 keV panel is the in-between-nuclei region used for count extraction (see text), and the red angular sector is the area used for the count extraction from the SE excess.

## 3.2 Image Reconstruction

Our analysis of the residual images after the subtraction of the nuclear AGN PSF models has demonstrated the presence of extra-nuclear X-ray emission in the central few hundred parsecs of NGC 6240, and has suggested energy-dependent differences in this emission. To further probe these differences, we have produced reconstructed images of the central region, using the image restoration algorithm Expectation through Markov Chain Monte Carlo (EMC2, Esch et al. 2004; Karovska et al. 2005, 2007, 2010; Wang et al. 2014; Fabbiano et al. 2018a, b; sometimes referred to as PSF deconvolution) with the PSF models described in Section 2.2.

EMC2 (Expectation through Markov Chain Monte Carlo) is a statistical deconvolution technique allowing reconstruction/restoration of an astronomical image affected by the blurring effects of the PSF (including telescope and detector effects). This technique uses a wavelet-like multiscale representation of the true image (unblurred by the PSF) in order to achieve smoothing at all scales of resolution simultaneously. Wavelet decomposition of the image allows the Poisson likelihood to be factored into separate parts, corresponding to the wavelet basis and each of these can be re-parametrized as a split of the intensity from the previous, coarser factor. A prior is assigned to these splits (which can be viewed as smoothing parameters) and a model is fit using Markov Chain Monte Carlo (MCMC) methods. This allows capturing small and large-scale structures in the image simultaneously, including diffuse emission as well as point-like sources (and sharp features in the image).This Bayesian model-based analysis is specifically applicable to low-counts Poisson data and allows assessment of the uncertainties in the reconstructed (deconvolved, restored) image. Details of this technique and examples are described in Esch et al. 2004.

To demonstrate how the diffuse emission recovered by the application of the EMC2 reconstruction to the data is not likely to be an artifact of the algorithm, we have simulated an image consisting of only two point-like sources, based on the two 6.0-6.4 keV model merged PSFs for the nuclear sources (Section 2). We have applied the same EMC2 reconstruction to this simulated image and to the real data. The results are shown in Figure 9. The EMC2 reconstructed image from the two PSF simulated image (left panel), shows only two point-like sources. Instead, the EMC2 reconstruction of the observed merged 6.0-6.4 keV image (right panel) shows clear extended emission, demonstrating that the extended emission cannot be an artifact of the reconstruction method. Similar simulations and thorough testing of the EMC2 techniques were done in the Esch et al. (2004; see e.g., their Figure 7).

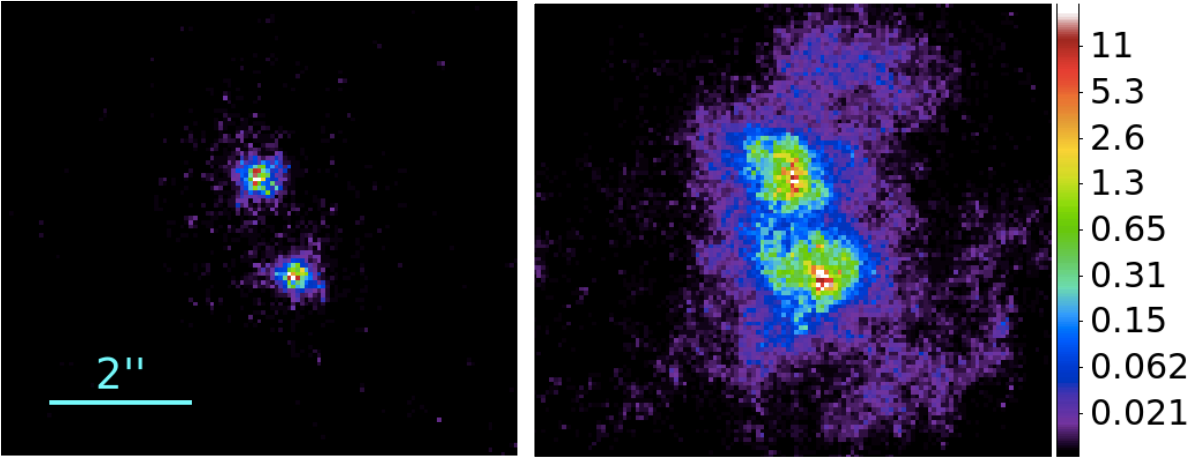

**Figure 9** – Left: EMC2 reconstructed image (100 iterations) of the 2 PSF model image. Right: EMC2 reconstruction of the observed merged 6.0-6.4 keV image. The images are with 1/8 pixel binning, displayed with the same logarithmic color scale in counts/pixel (on the right). N is to the top and E to the left.

Figure 10 shows the results of the 200 iterations EMC2 reconstruction in the three energy bands, which includes the larger spatial scales, for the 1/8 pixel images, with a further Gaussian smoothing with 2 pixel kernel. These images show diffuse emission extending out to ~ 5'' (~2.5 kpc) from the center (see also Wang et al. 2014). The 100 iterations reconstructions for the 1/16 pixel binned images of the inner region are shown in Figure 11, to emphasize the small scale features in the inner ~1 kpc region.

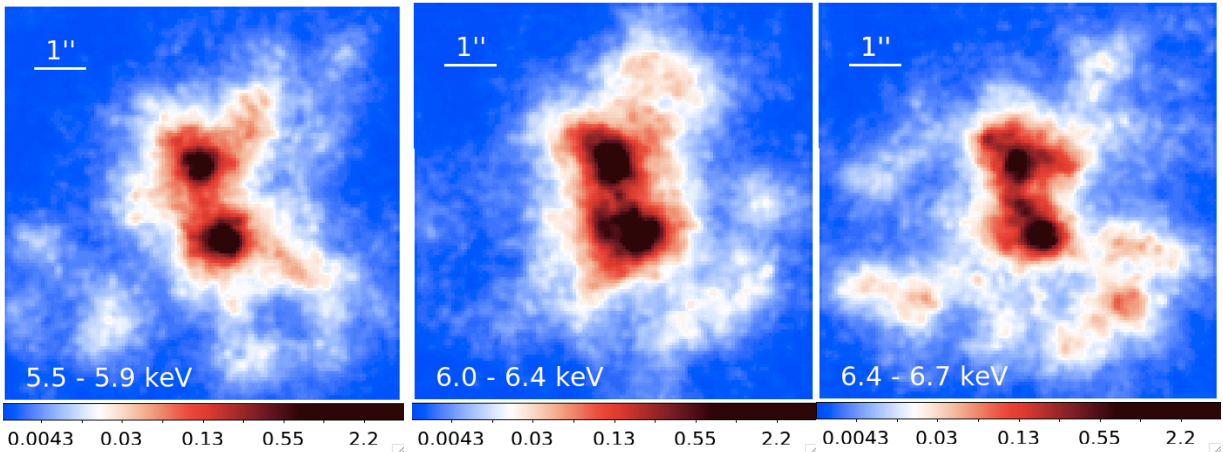

**Figure 10** – EMC2 reconstruction (200 iterations) of the X-ray emission in the central ~1 kpc of NGC 6240 (see text), in the hard 5.5-5.9 keV observed continuum, 6.0-6.4 keV (redshifted Fe I K$\alpha$) and 6.4-6.7 keV (redshifted Fe XXV) energy bands. The EMC2 images were further smoothed with a 2 pixel Gaussian kernel. N is to the top and E to the left in all the panels. The color scale is logarithmic in units of counts per image pixel (1/8 of the instrumental pixel). It has been stretched to emphasize the low surface brightness extended regions.

Figure 11 shows that there is extended X-ray emission both in the two nuclear regions and in between the nuclei. The differences between the images in the different energy bands suggests that these extended features are unlikely to arise from the PSF artifact, even in the regions where the artifact would contribute. The artifact is not included in the PSF model generated with the tools provided by the *Chandra* Data Center, so it would generate localized count excesses (up to ~6% of the PSF) in the areas within its footprint in Figure 11.

Figure 11 shows extended emission to the north of the southern AGN in both neutral and ionized Iron, in an area where the PSF anomaly is not expected to interfere. In the areas potentially affected by anomaly, there are spatial differences between the energy bands that suggest that some features may be real (Figure 12). These differences are also – albeit less strongly – reflected in the adaptively smoothed residuals in Figure 8. In particular, Figure 11 shows an apparent filament connecting the nuclei to the NE of the S AGN, in both hard continuum and Fe I K$\alpha$. Instead, a wider 'bridge' connects the two AGNs in the Fe XXV emission, to the W of the Fe I K$\alpha$ and continuum filaments (see lower-left panel of Figure 11). This bridge is statistically significant. Extracting the counts from this region from the raw Fe XXV image, we find an excess of 33 counts (4.2 $\sigma$) over the diffuse emission in adjacent regions. The bridge is evident also in the adaptively smoothed image (lower-right panel of Figure 11). While energy-dependent variability may in principle affect the images, it seems unlikely in our case. A point source, prominent in the 6.4-6.7 keV energy band, present in only a subset of the data should appear more localized than the 'bridge' emission. For example, see what the EMC2 reconstruction of point like emission shows in Fig. 9 (left). Moreover, hard continuum emission would also be expected from a thermal source that would produce a prominent Fe XXV line (Wang et al. 2014). No corresponding hard continuum feature is observed (Figure 12).

The Fe I K$\alpha$ image (top right of Figure 11) may suggest a possible third point-like emission region to the SE of the southern AGN, which is spatially consistent with the position of the 3$^{rd}$ possible nucleus of NGC 6240 reported by Kollatschny et al. (2020). The Fe XXV (bottom left) image does not show a source at the position of the third nucleus. There is an elongation in the SE direction in the hard continuum image (top left). We have shown that there is a prominent SE elongation in the residual images after PSF subtraction in all the energy bands, but especially in the continuum and Fe I K$\alpha$ (Figure 8; Section 3.1).

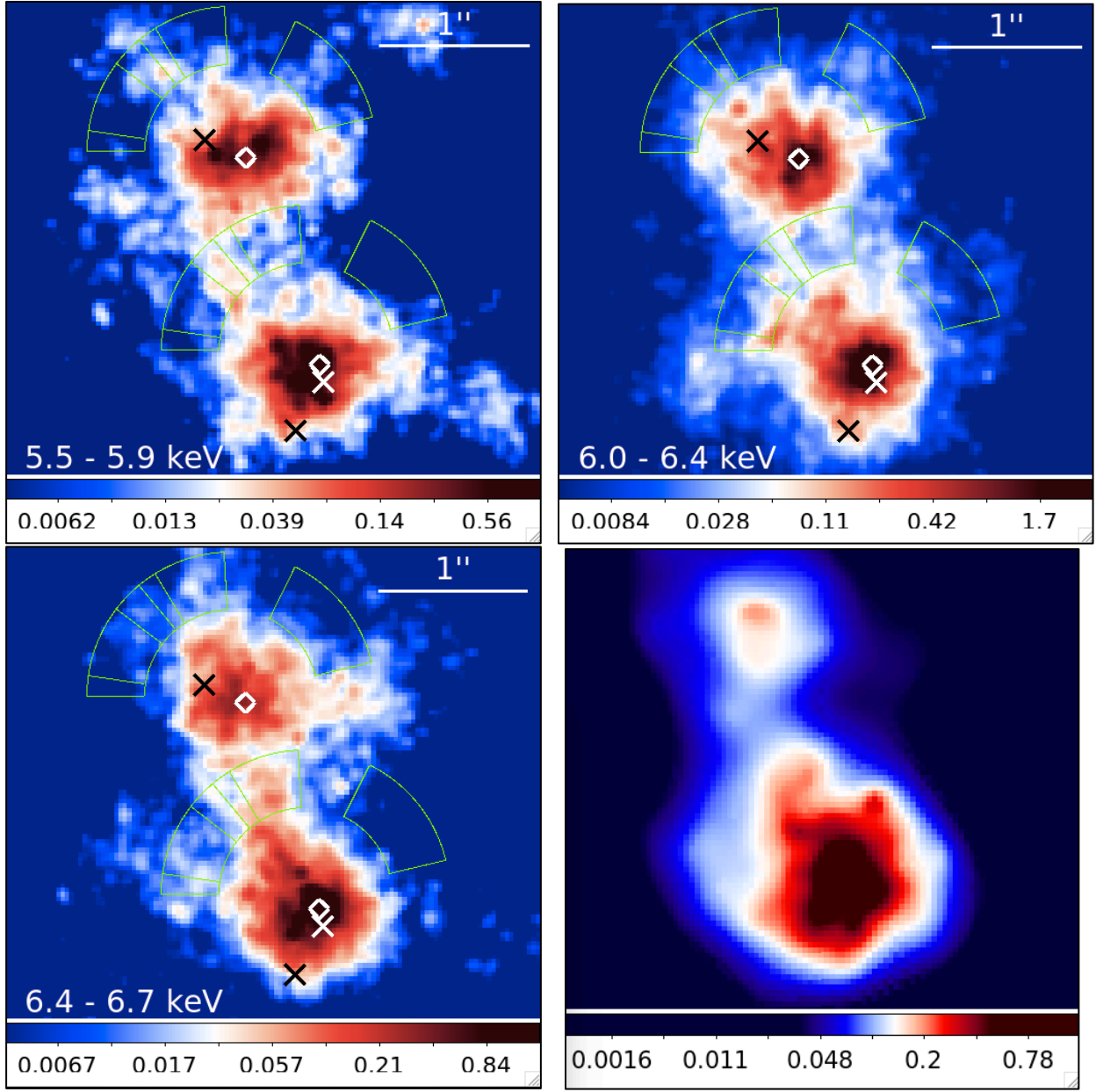

**Figure 11** – EMC2 reconstruction (100 iterations) of the X-ray emission in the central ~1 kpc of NGC 6240, in the indicated energy band. The EMC2 images were further smoothed with a 2 pixel Gaussian kernel. For comparison, the bottom right panel shows the adaptively smoothed image of the 6.4-6.7 keV band. The bridge between nuclear regions is evident also in this image. N is to the top and E to the left in all the panels. The color scale is logarithmic in units of counts per image pixel (1/16 of the instrumental pixel). The circular annuli are the regions contaminated by the PSF artifact (see Section 3.1). Diamonds mark the positions of the radio nuclei (Gallimore & Beswick 2004), while the crosses are at the position of the three optical nuclei (Kollatschny et al. 2020).

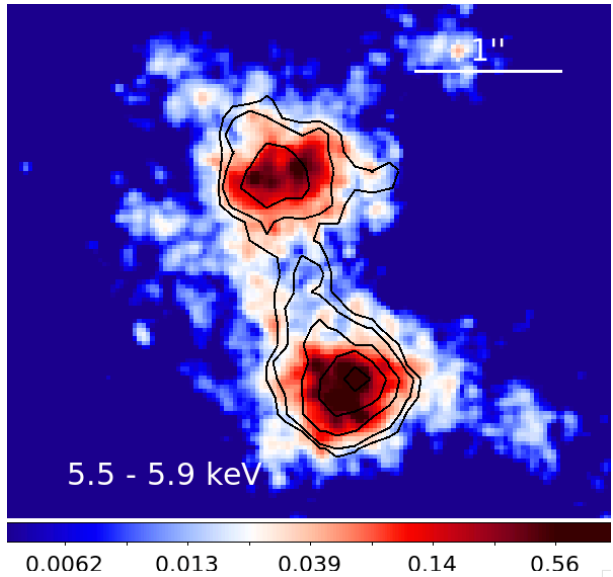

**Figure 12** – Contours of the EMC2 reconstructed Fe XXV intensity (lower-left panel of Figure 11) superimposed on the EMC2 reconstructed hard continuum map. No excess emission is observed in the continuum corresponding to the Fe XXV bridge. N is to the top and E to the left.

## 4. Discussion

Our high-resolution analysis of the spatial properties of the merger galaxy NGC 6240 (Section 3) has investigated the spatial properties of the extended Fe I K$\alpha$ (6.4 keV emitted), XXV (6.7 keV emitted) line and hard continuum (5.5-5.9 keV) emission in the central region of ~ 2 kpc radius that contains the two CT AGNs. Extended hard continuum and line X-ray emission in the central regions of NGC 6240 was previously reported by Wang et al. (2014), but its detailed morphology in the inner circum-nuclear regions was not investigated there.

By modeling and subtracting the *Chandra* PSF, we have established the presence of both Fe I and Fe XXV K$\alpha$ emission in the region between the two AGNs, at a significance of 5.6 $\sigma$ and 3.6 $\sigma$, respectively (Table 1). Moreover, we have discovered an emission feature apparently extending from the centroid of the southern AGN point-like emission to the SE out to ~1".5 (~770 pc). This feature is particularly prominent in the Fe I K$\alpha$ and hard continuum but is detected also in the Fe XXV band. This feature includes the region where a third nucleus of NGC 6240 has been claimed, based on high spatial and spectral resolution optical data (Kollatschny et al. 2020). The EMC2 image reconstruction confirms the presence of extended emission in the inner kiloparsec of NGC 6240. In particular, it suggests differences in the morphology of the emission in the region between the two AGNs, with the Fe XXV emission presenting a different spatial morphology than the continuum and Fe I K$\alpha$ line.

In what follows we discuss the implications of our results for the nature of the extended hard X-ray continuum and lines, the AGN-ISM interaction, and the active nuclei of NGC 6240, including the suggested 3rd nucleus (Kollatschny et al. 2020).

## 4.1 Thermal and Non-Thermal Extended X-ray Emission

Feruglio et al. 2013 and Wang et al. 2014 reported convincing evidence of thermal emission in the X-rays, thought to be caused by strong shocks in the ISM of NGC 6240. These thermal processes produce both hard continuum and Fe XXV line emission. The presence of extended Fe I K$\alpha$ emission, however, cannot be explained in this framework. While Fe XXV and the thermal hard continuum arise from the shocked interstellar medium (see Feruglio et al. 2013; Wang et al. 2014), Fe I K$\alpha$ is thought to arise from fluorescence, excited by hard nuclear X-ray photons interacting with dense clouds. Hard non-thermal continuum emission is also produced by the scattering of nuclear photons by these dense clouds (e.g. Baloković et al., 2018).

The presence of strong Fe I K$\alpha$ emission in CT AGNs suggested that this interaction occurred in the obscuring torus, a region of sub-parsec radius surrounding the AGN (Suganuma et al., 2006). However, more recent *Chandra* observations of several CT AGNs have detected extended Fe I K$\alpha$ emission and associated it with regions ranging from several tens of parsecs up to several kiloparsecs (Marinucci et al. 2012, 2017; Bauer et al. 2015; Koss et al. 2015; Fabbiano et al 2017, 2018a, b; Jones et al. 2020; Ma et al. 2020a). These observations suggest that fluorescence can also arise from the interaction of the AGN photons with dense molecular clouds in the host galaxy (Fabbiano 2018a), or in substructures, such as a circumnuclear disk (Marinucci et al 2012, 2017; Fabbiano 2019a). Even in the Milky Way, there is strong evidence of Fe I K$\alpha$ emitting clouds, first discovered with *ASCA* (Koyama et al., 1996) and more recently studied with *Chandra* and *XMM-Newton* (Ponti et al., 2015; Churazov et al. 2017a, b). In the CT AGNs studied so far, wherever extended Fe I K$\alpha$ is found, the hard continuum also presents similar extended emission components, consistent with scattering by dense non-nuclear clouds, not just the nuclear torus.

In ESO 428-G014, the first well-studied case of a CT AGN with hard continuum and Fe I K$\alpha$ emission extended out to a few kiloparsecs (Fabbiano et al. 2017, 2018a, b), the extended Fe I K$\alpha$ line emission has a large equivalent width (EW ~ 2 keV, Fabbiano et al 2017), typical of the large EW measured from the overall spectra of CT AGNs (Levenson et al. 2002). This large EW is consistent with the scattering/fluorescence picture that explains the CT AGN spectra (e.g. Baloković et al., 2018).

In NGC 6240, from the values in Table 1, we measure a smaller EW ~ 330±0.1 eV for the Fe I K$\alpha$ line of the extended component in the inner 2" (~1 kpc) circle. This EW is consistent with the value found by Wang et al. 2014, for a slightly larger emission area. This lower Fe I K$\alpha$ EW than typical of CT AGNs (~1 keV or more) could be explained by a mix of thermal and non-thermal continuum emission. We measure the non-thermal Fe I K$\alpha$ directly, as well as the continuum, which we can assume to be composed of non-thermal emission (from scattering of hard nuclear photons from the same dense clouds responsible for the Fe I K$\alpha$ line) and of thermal emission from shock-heated gas (see e.g., Wang et al 2014). If we assume that the EW of the extended component of the non-thermal Fe I K$\alpha$ line is a nominal EW~1 keV (e.g., Levenson et al 2002), from the definition of EW we can estimate the amount of extra continuum needed for decreasing the EW to the measured value. Under these assumptions, we derive that the non-thermal component could account for ~1/3 of the observed extended hard continuum emission in the central 2" radius circumnuclear region of NGC 6240. In the region between the two AGNs (the box in Figure 8), the non-thermal continuum emission may be less diluted. In this box we find that the Fe I K$\alpha$ EW~1.6±0.9 keV, consistent - within the large uncertainty - with a CT AGN Fe I K$\alpha$ EW, using the 'box excess' values in Table 1, which are corrected for the PSF artifact (Section 3.1).

## 4.2 AGN – ISM Interaction

Both Feruglio et al. 2013 and Wang et al. 2014 discussed the presence of hard shock-ionized ISM emission from the central regions of NGC 6240. In Feruglio et al.'s paper, this shock ionization was related to the strong nuclear outflows (v ~ 400-800 km s$^{-1}$) suggested by the CO kinematics. Wang et al. instead suggested stellar winds from the active star-forming region in the center of this galaxy, as a cause. Their star-formation ionization picture was supported by the similarity of the Fe XXV spatial distribution with that of the near IR H$_2$(1-0) 2.12 µm emission from Max et al. (2005). However, a more recent paper by Cicone et al. (2018), based on *ALMA* observations, reports a large wide-spread outflow from NGC 6240, whose energetics challenge the star-formation only powered hypothesis. In a study based on *HST* and *VLT/SINFONI* observations, Müller-Sánchez et al. (2018) conclude that both nuclear and starburst outflows are present in NGC 6240.

The presence of extended Fe I K$\alpha$ emission provides another view of the AGN-ISM interaction, that of hard X-ray photons interacting with dense molecular clouds. As already noted for the Milky Way and ESO 428-G014 (Koyama et al. 1996; Ponti et al. 2015; Churazov et al. 2017a, b; Fabbiano et al. 2018a, 2019a), high spatial resolution observations of the Fe I K$\alpha$ emission may provide a complementary way to direct CO observations, for mapping giant molecular clouds complexes in galaxies.

We have established that there is a highly significant excess of Fe I K$\alpha$ emission in the region between the two AGNs of NGC 6240 (Table 1), although we cannot map this region in detail because of the intervening *Chandra* PSF artifact (see Section 3.1). The EMC2 reconstructed maps (Section 3.2) tentatively suggest possible filaments connecting the two AGNs, both in Fe I K$\alpha$ and hard continuum, that appear displaced from the Fe XXV bridge that joins the two nuclei (Figure 12). While the Fe XXV emission is thermal and believed to originate from shock-ionized hot ISM regions (Wang et al. 2014), the inter-nuclear continuum and Fe I K$\alpha$ emission is likely to be the result of the scattering and fluorescence of the AGN photons interacting with the molecular clouds.

Manohar & Scoville (2017) imaged the central region of NGC 6240 in the 3 mm band transitions of CO, HCN, HCO$^+$, HNC, and CS, finding that the line ratios are not consistent with the chemistry of 'X-ray Dominated Regions' (AGN excitation). The concentration of CO (2-1) emission in region between the nuclei (Treister et al. 2020) appears to be consistent with this conclusion. Comparison of *Chandra* and *ALMA* observations have shown an anticorrelation near the active nuclei of ESO 428-G014 (Feruglio et al. 2020) and NGC 2110 (Fabbiano et al 2019b), suggestive of X-ray excitation of the molecular clouds - a localized lack of CO (2-1) emission in regions of prominent X-ray emission near the AGN - which is not present here. The situation of NGC 6240 is instead suggestive of fluorescent emission of the dense molecular clouds in the inter-nuclei regions, stimulated by the interaction with nuclear X-ray photons, as reported for the extended components of CT AGNs imaged with *Chandra*. A similar example of near-nucleus interaction is found in NGC 5643, where a Fe I K$\alpha$ spatially linear emission feature has been associated with the outer regions (Fabbiano et al. 2018c), of the rotating nuclear molecular disk, detected with *ALMA* (Alonso-Herrero et al. 2018).

However, some localized X-ray suppression of the CO (2-1) emission may be present, if we take the EMC2 maps (Figures 11, 12) at face value and we compare them with the high angular resolution CO (2-1) *ALMA* images in the recent paper by Treister et al. (2020). These authors resolved a system of clouds in the region between the two AGNs and studied their kinematics and velocity dispersion, concluding that there is a high-velocity turbulent stream of molecular clouds in between the two nuclei, resulting from the interaction of the two circumnuclear disks. Figure 13 shows the EMC2 Fe XXV contours (as in Figure 11) superimposed on the CO (2-1) intensity map of Treister et al. (2020; their Figure 1). The Fe XXV line emission appears displaced from the peak of the CO (2-1) emission, suggesting a possible local X-ray excitation and suppression of the CO (2-1) emission. We cannot however exclude that the hot shock-ionized gas, responsible for the Fe XXV emission is physically displaced by the densest molecular clouds detected with *ALMA*.

The region to the SE of the southern AGN (6240S in Figure 12; Treister et al. 2020, their Fig. 1), where we find evidence of extended emission, particularly prominent in the Fe I K$\alpha$ line and hard continuum (Section 3.1, Figure 8), is also rich of CO emitting clouds, which may belong to a rotating disk associated with the S AGN (Treister et al. 2020). In this region we find a fairly low EW~340 eV for the Fe I K$\alpha$ line, suggesting a mix of thermal and scattered emission for the hard continuum. The presence of correlated Fe XXV emission is consistent with shock-excitation in this region (Wang et al. 2014).

**4.3 The CT AGNs of NGC 6240**

With *Chandra*, Komossa et al. (2003) discovered the double CT AGN system of NGC 6240. Given the complexity of the hard circumnuclear emission, the spectra extracted (r< 1" for each AGN) by Komossa et al. for their spectral analysis are contaminated by the diffuse emission in the area (see Nardini 2017). Using the 1/16 pixel binned data, we can estimate directly from our spectral imaging the EW of the Fe I K$\alpha$ line within 0".2 (~100 pc) of the X-ray centroid of each AGN. For the N AGN, we find EW=0.9±0.2 keV, consistent with the typical values for CT AGNs (Levenson et al., 2002) For the S AGN instead we derive a lower EW=0.47±0.1 keV, suggesting that the continuum may be augmented by non-nuclear thermal emission. Komossa et al., (2003) do not quote EW values; their plotted spectra appear consistent with the values found here. Even at these small radii, however, the SE excess may affect the data (Figures 6 and 8).

It is worth pointing out that with Chandra we can explore down to these small radii, which are within the sphere of influence of each nuclear black hole, ~235-212 pc (Medling et al. 2015; see Figure 7), i.e. the regions within which the black hole contributes >50% of the total mass (although these figures have been somewhat diminished by more recent measurements of molecular gas in the areas; Medling et al. 2019). Treister et al. (2020) report that the S nucleus has peculiar kinematics within its sphere of influence (~0.4''), the same radius within which we find a discrepancy between the Chandra PSF and the data (SE excess). Medling et al. (2019) also report a larger contribution of molecular gas mass in the S AGN area. The SE excess is also in the direction of the third nucleus reported by Kollatschny et al. (2020), but the distribution of residuals (Figure 8) suggests that this excess is not due to a single point source.

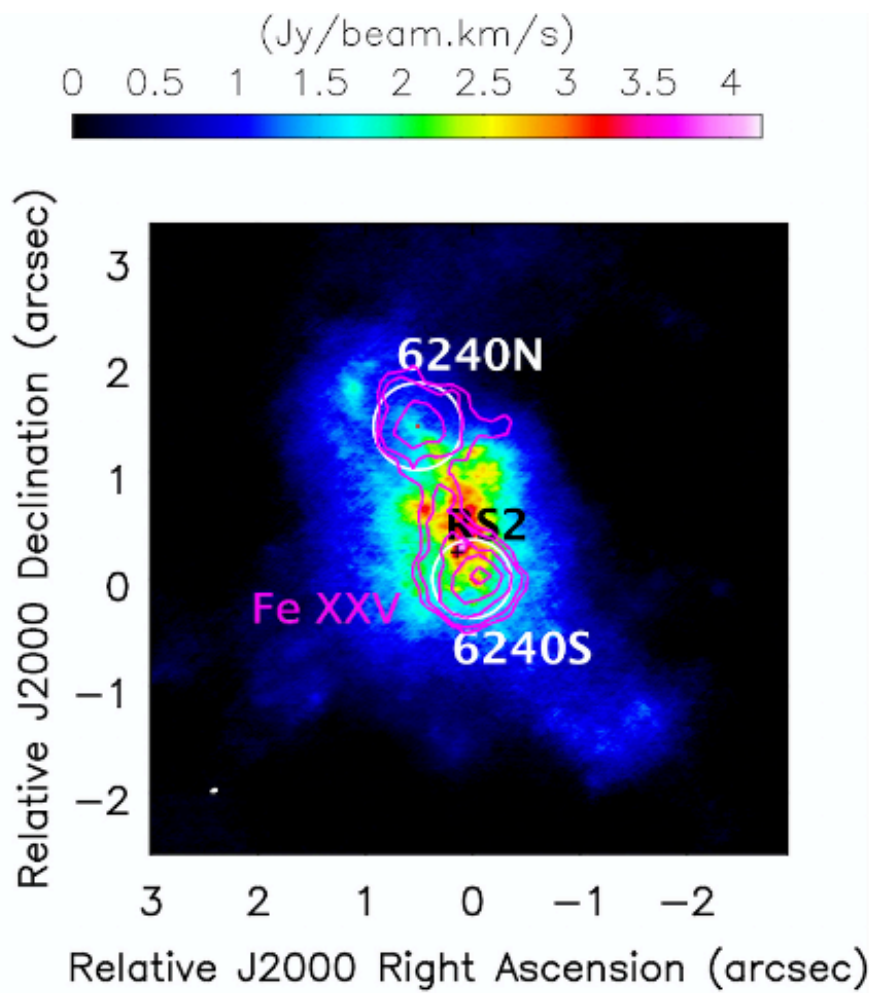

**Figure 13** – Integrated CO (2-1) intensity map from Trester et al. 2020 with overlayed Fe XXV contours from the EMC2 image (magenta; see Figs. 10-11). The two white circles represent the spheres of influence of the two AGNs, RS2 marks the position of the homonymous supernova (see Treister et al. 2020). The *Chandra* contours were overlayed by assuming that the X-ray centroids are coincident with the two nuclear BHs at the center of the spheres of influence.

Kollatschny et al., (2020) find that the N AGN, the S AGN and the possible 3$^{rd}$ AGN all lie close to the LINER/AGN boundary in the BPT excitation diagram (Veilleux & Osterbrock 1987). This suggests that all three are AGNs. However, BPT mapping of other nearby AGNs shows that regions of LINER emission are often found surrounding extended regions of AGN emission on a similar scale of hundreds of parsecs, with boundary values found in the interface between them (Maksym et al., 2016; Ma et al., 2020b) Hence the excitation diagnostic alone is not enough to require a 3$^{rd}$ nucleus.

The EMC2 image reconstruction of the Fe I K$\alpha$ emission (top right panel of Figure 11) suggests a possible source at the 3$^{rd}$ nucleus position. If we conservatively use the residual map (Section 3.1) and extract the counts from a circle of 0".2 radius at the position of the reported 3$^{rd}$ nucleus we obtain ~30±5.9 counts. For the same circular region, we estimate an Fe I K$\alpha$ EW=400±200 eV. Given the possible dilution of the continuum by thermal emission, Fe I K$\alpha$ is a more reliable measure of the AGN luminosity. The 3$^{rd}$ nucleus has ~1/4 of the counts detected from a similar circle at the centroid of the N AGN from the raw data in the same energy band. Assuming the N AGN to be a typical CT AGN benchmark, given its large Fe I K$\alpha$ EW (see above) and scaling by the X-ray luminosity of the N AGN (Komossa et al. 2003) by the same factor, we estimate an X-ray luminosity of ~ 2 x 10$^{41}$ erg s$^{-1}$ for the third nucleus. Although this is not a detection, it is interesting to notice that the mass reported for the third nucleus (Kollatschny et al. 2020) is ~1/4 of that of the N AGN. Alternatively, we can use the EMC2 Fe I K$\alpha$ image (Figure 11), which has 9 counts from a 0".2 circle encompassing the region coincident with the 3$^{rd}$ nucleus position. This reconstruction was run with 100 iterations to be more sensitive to small scale features, so it is likely to underestimate the total counts in the observed image, which also would include larger-spatial-scale extended emission. This would result in a 3s upper limits of 18 counts on the source counts and 1.4 x 10$^{41}$ erg s$^{-1}$ upper limit on the luminosity.

## 5. Summary and Conclusions

We have reanalyzed the cumulative ACIS S *Chandra* data set pointed at the double AGN of the NGC 6240 merging galaxy, focusing on the hard energy bands containing the hard spectral continuum (5.5-5.9 keV), the redshifted Fe I K$\alpha$ line (6.0-6.4 keV), and the redshifted Fe XXV line (6.4-6.7 keV). We have exploited to the full the angular resolution of the *Chandra* telescope, by using subpixel binned imaging (1/8 and 1/16 of the ACIS pixel of 0".492). We have also modeled the *Chandra* PSF by comparing pre-flight calibration model to the data for the two bright AGNs. We have used two complementary approaches to the analysis: (1) studying the residuals after PSF subtraction in the three energy bands, and (2) producing reconstructed EMC2 images to correct for the PSF in the observed images.

We are able to resolve structures extending from ~1 kpc to smaller scales near both nuclei. In the S AGN, we find a <200 pc structure, within the sphere of influence of this BH, which may be connected with more extended emission in both continuum and Fe lines in the ~2" (~1 kpc) region surrounding the nuclei, in the region between the N and S AGN, and in a sector of PA 120-210 deg. extending to the SE from the centroid of the S AGN surface brightness.

The extended Fe I K$\alpha$ emission cannot be explained by the thermal processes that were invoked to explain the extended hard continuum and Fe XXV emission (Wang et al. 2014; Feruglio et al.

2013). Fe I Kα is instead likely to originate from fluorescence of X-ray photons interacting with dense molecular clouds (as found in other CT AGNs, e.g. ESO 428-G014, Fabbiano et al. 2017).

Comparison of the Fe I Kα EW in different areas with the nominal EW~1 keV expected from CT AGNs, and also the extended emission of ESO 428-G014, suggests that:
(1) in the circumnuclear ~1 kpc radius region non-thermal processes are responsible for ~1/3 of the hard continuum emission, while
(2) in the region in-between the two AGN the non-thermal contribution is higher.

Extracting counts within 100 pc (0".2) from the centroid, we find that:
(3) the EW of the N AGN is 0.9±0.2 keV, consistent with the typical values for CT AGNs (Levenson et al., 2002),
(4) the EW of the S AGN is lower EW=0.47±0.1 keV, suggesting that the continuum may be augmented by non-nuclear thermal emission. This thermal emission may be connected with the 'SE excess' observed in this source.

The EMC2 reconstructed images suggest possible filaments in the region between the two AGNs. Particularly significant is a bridge of Fe XXV emission connecting the nuclei. This bridge appears displaced from the turbulent CO (2-1) stream found by Treister et al. (2020) with *ALMA*, suggesting a displacement of hot shocked gas and molecular clouds and/or depletion of the CO emission in the area because of X-ray excitation.

We do not find strong evidence of X-ray emission associated with the 3$^{rd}$ nucleus proposed by Kollatschny et al. (2020), although this region is associated with the SE excess, and a possible clump of Fe I Kα emission is suggested in this area by the EMC2 reconstruction. We estimate a ~1.4 – 2 10$^{41}$ erg s$^{-1}$ upper limit on the luminosity of a third nucleus.

Our results show the importance of comparing *ALMA* and optical data with similarly high angular resolution X-ray data. A future high-resolution, larger-area X-ray telescope will be needed to pursue further these studies in the post-*Chandra* future.

We retrieved data from the NASA-IPAC Extragalactic Database (NED) and the *Chandra* Data Archive. For the data analysis, we used the CIAO toolbox and DS9, developed by the *Chandra* X-ray Center (CXC). This work was partially supported by the NASA contract NAS8-03060 (CXC). JW acknowledges support by the National Key R&D Program of China (2016YFA0400702) and the NSFC grants (U1831205, 11473021, 11522323). EN is supported by ASI-INAF grant No. 2017-14-H.0.